\documentclass[trackchanges]{aastex701}

\begin{document}

\title{Quasiperiodic Slipping Motion of Flare Ribbon Fine Structures Anchored in a Sunspot Light Bridge}

\author[0009-0001-6754-8543]{Tianyuan Chen}
\affiliation{Yunnan Observatories, Chinese Academy of Sciences, Kunming 650216, People's Republic of China}
\affiliation{University of Chinese Academy of Sciences, Beijing 100049, People's Republic of China}
\email{chentianyuan@ynao.ac.cn}

\author[0000-0003-2891-6267]{Xiaoli Yan}
\affiliation{Yunnan Observatories, Chinese Academy of Sciences, Kunming 650216, People's Republic of China}
\affiliation{Yunnan Key Laboratory of Solar Physics and Space Science, Kunming 650216, People's Republic of China}
\email{yanxl@ynao.ac.cn}

\correspondingauthor{Xiaoli Yan}
\email{yanxl@ynao.ac.cn}

\author[0000-0002-6526-5363]{Zhike Xue}
\affiliation{Yunnan Observatories, Chinese Academy of Sciences, Kunming 650216, People's Republic of China}
\affiliation{Yunnan Key Laboratory of Solar Physics and Space Science, Kunming 650216, People's Republic of China}
\email{zkxue@ynao.ac.cn}

\author[0000-0003-4393-9731]{Jincheng Wang}
\affiliation{Yunnan Observatories, Chinese Academy of Sciences, Kunming 650216, People's Republic of China}
\affiliation{Yunnan Key Laboratory of Solar Physics and Space Science, Kunming 650216, People's Republic of China}
\email{wangjincheng@ynao.ac.cn}

\author[0000-0002-9121-9686]{Zhe Xu}
\affiliation{Yunnan Observatories, Chinese Academy of Sciences, Kunming 650216, People's Republic of China}
\affiliation{Yunnan Key Laboratory of Solar Physics and Space Science, Kunming 650216, People's Republic of China}
\email{xuzhe6249@ynao.ac.cn}

\author[0000-0003-0236-2243]{Liheng Yang}
\affiliation{Yunnan Observatories, Chinese Academy of Sciences, Kunming 650216, People's Republic of China}
\affiliation{Yunnan Key Laboratory of Solar Physics and Space Science, Kunming 650216, People's Republic of China}
\email{yangliheng@ynao.ac.cn}

\author[0000-0001-9491-699X]{Yadan Duan}
\affiliation{Yunnan Observatories, Chinese Academy of Sciences, Kunming 650216, People's Republic of China}
\affiliation{Yunnan Key Laboratory of Solar Physics and Space Science, Kunming 650216, People's Republic of China}
\email{duanyadan@ynao.ac.cn}

\author[0000-0002-0464-6760]{Yian Zhou}
\affiliation{Yunnan Observatories, Chinese Academy of Sciences, Kunming 650216, People's Republic of China}
\email{zhouyian@ynao.ac.cn}

\author[0009-0003-9377-1989]{Zongyin Wu}
\affiliation{Yunnan Observatories, Chinese Academy of Sciences, Kunming 650216, People's Republic of China}
\affiliation{University of Chinese Academy of Sciences, Beijing 100049, People's Republic of China}
\email{wuzongyin@ynao.ac.cn}

\author{Qifan Dong}
\affiliation{Yunnan Observatories, Chinese Academy of Sciences, Kunming 650216, People's Republic of China}
\affiliation{University of Chinese Academy of Sciences, Beijing 100049, People's Republic of China}
\email{dongqifan@ynao.ac.cn}

\author{Guotang Wu}
\affiliation{Yunnan Observatories, Chinese Academy of Sciences, Kunming 650216, People's Republic of China}
\affiliation{University of Chinese Academy of Sciences, Beijing 100049, People's Republic of China}
\email{wuguotang@ynao.ac.cn}

\author{Xinsheng Zhang}
\affiliation{Yunnan Observatories, Chinese Academy of Sciences, Kunming 650216, People's Republic of China}
\affiliation{University of Chinese Academy of Sciences, Beijing 100049, People's Republic of China}
\email{zhangxinsheng@ynao.ac.cn}

\begin{abstract}
We used high-resolution observations from the New Vacuum Solar Telescope  and the Solar Dynamics Observatory  to perform a detailed multiwavelength analysis of the fine structures in the flare ribbon of a C3.9-class flare on 2021 April 22. A segment of the flare ribbon was rooted in a sunspot light bridge and exhibited discrete sub-structures that we term ``burrs,"  with equivalent diameters of 233--895 km and inter-core separations of 1129--1739 km. These structures are characterized by discrete redshifted cores, accompanied by  ``tails" (length 700--1370 km, width 310--600 km) exhibiting faint blueshifts. These structures exhibit systematic slipping motions along the ribbon, with apparent velocities decelerating from about 40--21 km s$^{-1}$, and display a distinct quasiperiodicity of $\sim$6 minutes in H$\alpha$ and EUV passbands. Differential Emission Measure (DEM) analysis confirms the emitting plasma is multi-thermal, dominated by temperatures of 1–-2 MK. The observed morphology and kinematics  are consistent with the scenario of impulsive energy deposition by precipitating plasmoids (oblique flux ropes) originating from tearing-mode fragmentation in the coronal current sheet. The specific spatiotemporal correlation between the tails and blueshifts supports the hypothesis of untwisting magnetic flux ropes. Furthermore, the $\sim$6 minute periodicity suggests that the reconnection process may be modulated by photospheric $p$-mode oscillations coupled with the tearing-mode instability. Our findings provide observational evidence that these light bridge anchored fine structures constitute elementary units of flare energy release.

\end{abstract}



\section{Introduction} \label{sec:Introduction}

 Solar flares are among the most energetic phenomena in the solar system, capable of releasing up to $10^{32}$  erg of energy within a few minutes \citep{Fletcher_2011}. This energy is released from the coronal magnetic field by magnetic reconnection \citep{Parker_1957,Kopp_1976,Zweibel_2009}. A fraction of the released energy may drive the outward ejection of coronal plasma as coronal mass ejections (CMEs), which are the main driver of disastrous space weather \citep{Bothmer_1994,Forbes_2006,Shibata_2011}. Another fraction of the energy released by reconnection is thought to be transported downward along newly reconnected magnetic field lines. When this energy is deposited in the dense chromosphere, it is rapidly thermalized, producing strong local plasma heating and intense emission across multiple electromagnetic bands, which are observed as visually bright flare ribbons \citep{Fletcher_2011,Benz_2017}.

Recent high-resolution observations reveal that flare ribbons are not uniform bright bands but are composed of a series of complex fine structures. These include compact flare kernels \citep{Xu_2012,Graham_2015,Pietrow_2024,Lörinčík_2025,Zhang_2025}, elongated, jet-like plasma columns known as ``riblets" \citep{Thoen_2026,Singh_2025},  spatially quasiperiodic dotlike structures \citep{Thoen_2025a,Yadav_2025}, wavelike or spiral-shaped structures \citep{Ofman_2011,Brannon_2015,Parker_2017}. Furthermore, these structures are often observed to move rapidly along the flare ribbons \citep{Li_2015,Lörinčík_2025}. Neither these complex morphologies nor their pronounced dynamic behaviors can be explained by the classical 2D CSHKP model \citep{Carmichael_1964,Sturrock_1966,Hirayama_1974,Kopp_1976}. Therefore, understanding these phenomena inevitably requires the introduction of more sophisticated, intrinsic 3D processes.

Regarding the origin of these fine structures, several structural models have been proposed. The tearing-mode instability in the flare current sheet posits that an extended current sheet will spontaneously fragment into a sequence of discrete plasmoids and current filaments during the reconnection process \citep{Shibata_2001}. The analytical model by \citet{Wyper_2021}  shows that oblique tearing modes in the current sheet can directly generate wavelike and spiral fine structures in both the hook and straight segments of flare ribbons, with morphologies that are remarkably similar to observations. Recent observations from the latest generation of high-resolution telescopes have provided strong support for this theoretical picture: \citet{Thoen_2025a} and \citet{Yadav_2025}, using the Swedish Solar Telescope (SST; \citealt{Scharmer_2003}) and the Daniel K. Inouye Solar Telescope (DKIST; \citealt{Rimmele_2020}), respectively, directly resolved quasiperiodically arranged discrete brightenings along flare ribbons, confirming that these discrete kernels are the chromospheric manifestation of the tearing-mode instability and constitute the fundamental units of flare energy release. However, beyond this mainstream model, the morphology of these fine structures is also jointly shaped by downstream effects of reconnection and the complex response of the lower atmosphere. For example, fast reconnection downflows that impact the loop-top of flare loops can produce a termination shock whose disturbances propagate along the loop to the footpoints and cause localized, transient brightenings \citep{Forbes_1986,Chen_2019,Samanta_2021}. Meanwhile, the chromosphere’s nonlinear response to energy injection, such as intense evaporation  condensation cycles \citep{Fisher_1985,Graham_2015} and Joule heating in partially ionized conditions \citep{Sharykin_2014} can generate or modify rapidly evolving fine brightenings.

Slipping magnetic reconnection describes the continuous reconnection of magnetic field lines along quasi-separatrix  layers (QSLs; \citealt{Priest_1995,Demoulin_1996}), which causes their chromospheric footpoints to appear to systematically slip along flare ribbons \citep{Aulanier_2005}. Numerical studies demonstrated that such reconnection naturally leads to systematic slippage and, in the slip-running regime, to very rapid (super-Alfvénic) apparent motions \citep{Aulanier_2006}. \citet{Dudík_2014} provided compelling high-resolution observations of apparently slipping flare loops with sub-Alfvénic speeds. \citet{Li_2015} and \citet{Li_2018}  reported quasiperiodic slipping patterns in eruptive and confined flares, respectively, suggesting that the reconnection process may be modulated by magnetohydrodynamic (MHD) waves. Recently, \citet{Lörinčík_2025} used high-cadence Interface Region Imaging Spectrograph (IRIS;\ \citealt{Pontieu_2014}) high-cadence  slit data ($\sim$2 s) to reveal flare kernel motions at apparent speeds of order $10^3$ \  km s$^{-1}$ , providing the direct observational evidence for slip-running (super-Alfvénic) reconnection in QSLs.

Recent spectroscopic analysis \citep{Pietrow_2024} has identified multiple fine structures within flare ribbons with distinct physical properties, and specifically noted that when flares occur above complex magnetic regions containing sunspot light bridges, the intricate magnetic boundaries in the underlying photosphere provide crucial constraints on the fine-scale structures and the diverse modes of energy dissipation in the overlying atmosphere.

Research on fine structures in flare ribbons has increased, but their physical origins remain unclear, particularly because detailed  observational investigations of the dynamic evolution of individual fine structures and their coupling to the underlying  magnetic field are lacking. In this work,  we employed high spatial and temporal resolution observations from the New Vacuum Solar Telescope (NVST; \citealt{Liu_2014}) and the Solar Dynamics Observatory (SDO; \citealt{Pesnell_2012}) to perform a detailed analysis of the fine structures of the flare ribbons in a C3.9-class flare on 2021 April 22. Our observations, for the first time, revealed a series of transient, small-scale burr structures that exhibited clear slipping motion patterns and were anchored in the complex magnetic topology of a sunspot light bridge. This paper is organized as follows: In Section~\ref{sec:Observation and Data Analysis} we introduce the equipment used for observation and the methods for processing data. In Section~\ref{sec:Result} we describe the results of the observation. Finally, in Section~\ref{sec:Discussion}, we summarize our findings and discuss the comparison of spatial scales between burrs and other fine structures observed under high-resolution conditions, the potential physical mechanisms behind the sub-Alfvénic slipping motion of the burrs, the quasiperiodic nature of the fragmented magnetic reconnection, and the influence of the sunspot light bridge's magnetic topology on these fine structures.


\begin{figure*}[t!]
\includegraphics[trim={0cm 0cm 0cm 0cm}, clip, width=\textwidth]{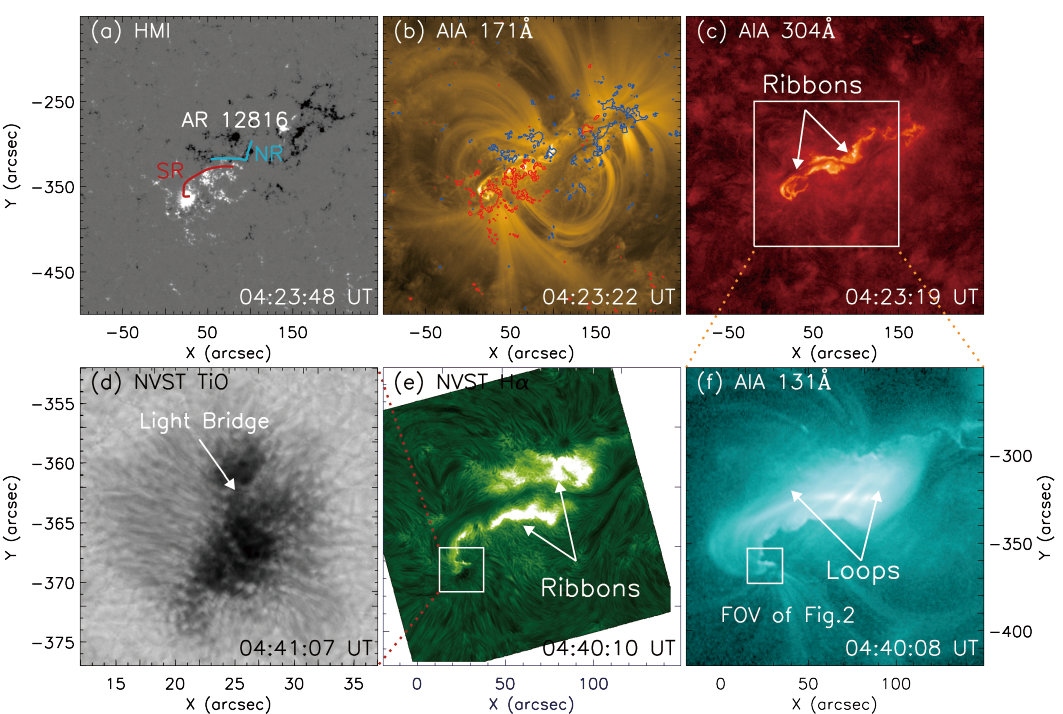}
\caption{
\label{fig:1}Multiwavelength overview of the C3.9-class flare in AR 12816 on 2021 April 22. Panel (a): SDO/HMI LOS magnetogram showing the magnetic field distribution of the active region. The red and blue curves delineate the J-shaped positive-polarity ribbon  and the L-shaped negative-polarity ribbon , respectively. Panel (b): SDO/AIA 171 \AA\  image showing the brightened coronal loop structures during the flare. Panel (c): SDO/AIA 304 \AA\  image clearly outlining the flare ribbons. Panel (d): NVST TiO image showing the fine optical structure of the main sunspot at the footpoint of the SR. Panel (e): NVST H$\alpha$ image displaying the corresponding chromospheric structure of the flare ribbons. Panel (f): SDO/AIA 131 \AA\  image showing the post-flare loops formed in the later phase of the flare.}
\end{figure*}

\section{Observation and Data Analysis} \label{sec:Observation and Data Analysis}

In this work, we investigate the C3.9-class solar flare that took place in NOAA Active Region (AR) 12816 on 2021 April 22. The analysis is based on a coordinated dataset from multiple instruments, including the ground-based NVST and the space-based SDO. The NVST provides high-resolution imaging observations of the photosphere and chromosphere \citep{Liu_2014,Yan_2020}, including photospheric TiO 7058 \AA\  images with a cadence of 30 s and chromospheric H$\alpha$ line-center (6562.8 \AA) and off-band (6562.8 $\pm $\ 0.4 \AA) images with a cadence of 45 s. The pixel scale is $ 0\farcs 165 $ after image reconstruction. The Atmospheric Imaging Assembly (AIA; \citealt{Lemen_2012}) on board the SDO provides multiwavelength images in several ultraviolet (UV) and extreme-ultraviolet (EUV) passbands. For this study, we utilize images from seven of the EUV channels: 94, 131, 171, 193, 211, 304, and 335 \AA. These images provide a cadence of 12 s and a spatial resolution of $1\farcs 5$ (pixel size of $0\farcs 6$), allowing for detailed tracking of the burrs evolution during the flare. The Helioseismic and Magnetic Imager (HMI; \citealt{Scherrer_2012}) on board the SDO also provided crucial data for this study. HMI continuously observes the photospheric magnetic field with unprecedented precision and stability, generating full disk line-of-sight (LOS) magnetograms every 45 s with a pixel resolution of $ 0\farcs 5 $. This capability allowed us to characterize the detailed structure and dynamic evolution of the magnetic field in the flaring region.

To investigate the LOS dynamical evolution of the burrs, we constructed Doppler proxy  maps using the equation:   $ D = (B - R) / (B + R) $ \citep{Langangen_2008,Awasthi_2019}. In this formula, B and R are the intensities of the H$\alpha$ blue-wing (-- 0.4\ \AA ) and red-wing (+ 0.4 \AA ) images, respectively. The resulting Doppler proxy maps are thus a proxy for the LOS velocity and do not provide an absolute Doppler velocity.

\begin{figure*}[t!]
\includegraphics[trim={0cm 0cm 0cm 0cm}, clip, width=\textwidth]{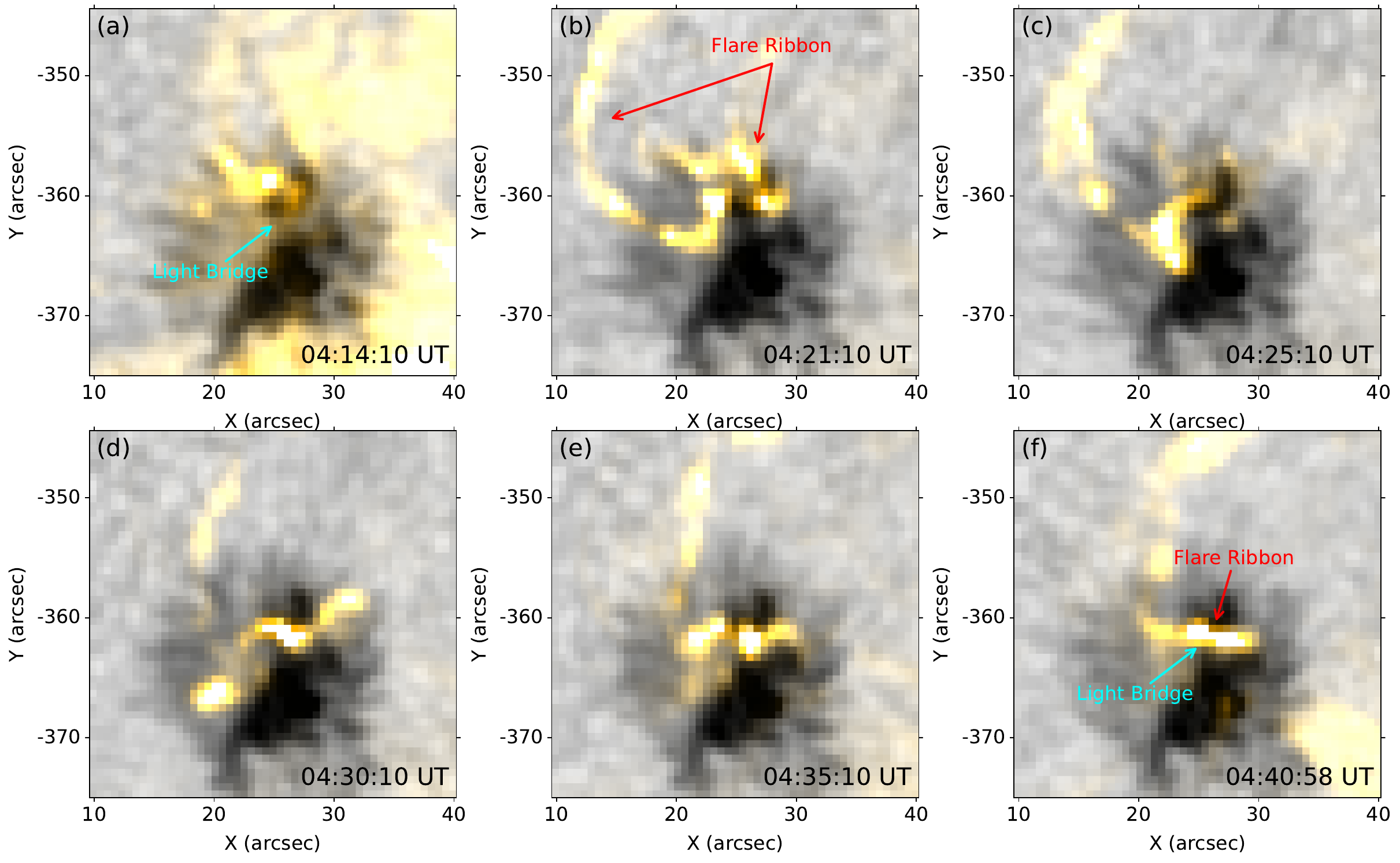}
\caption{
\label{fig:2}Dynamic evolution of the flare ribbon footpoint in the sunspot. This sequence of co-aligned AIA 171 \AA\  images and HMI continuum intensity images displays the continuous evolution of the flare ribbon footpoint near the sunspot from 04:14:10 UT to 04:40:58 UT. The images reveal the process from an initial curved bright arc (panels (a) and (b)) to several discrete bright kernels (panels (c)–(e)), which finally stabilize at the location of a light bridge (panel (f)).}
\end{figure*}

We performed a DEM analysis \citep{Cheung_2015,Su_2018}  to diagnose the thermal properties of the plasma in the flare burrs. Using near-simultaneous observations from six AIA channels (94, 131, 171, 193, 211, and 335 \AA), a sparse inversion technique was applied to each pixel to solve for its DEM function, which describes the plasma's temperature distribution along the LOS. From the resulting DEM curves, we derived key macroscopic quantities by integration and weighted averaging: the DEM weighted average temperature for bulk thermal characterization, the total emission measure (EM) for total radiative output, and the emission measure in specific temperature bins ($\log_{10} T$ = 5.7--6.0, 6.0--6.3, and 6.3--6.6) to investigate the spatial distribution of different temperature components.

Adopting the thresholding approach described by \citet{Thoen_2025a}, we applied a full-width at half-maximum (FWHM) criterion to the H$\alpha$ red-wing images acquired at +0.4 \AA\  to delineate the boundaries of the fine-scale structures. The background reference was defined as the median intensity of the entire field of view (FOV), a statistic chosen to mitigate biases from extreme flaring pixels. The threshold for each structure was determined as the midpoint between the local peak intensity and this background level. The spatial scale of each identified feature was subsequently characterized by the equivalent diameter derived from its projected area.

\begin{figure*}[t!]
\includegraphics[trim={0cm 0cm 0cm 0cm}, clip, width=\textwidth]{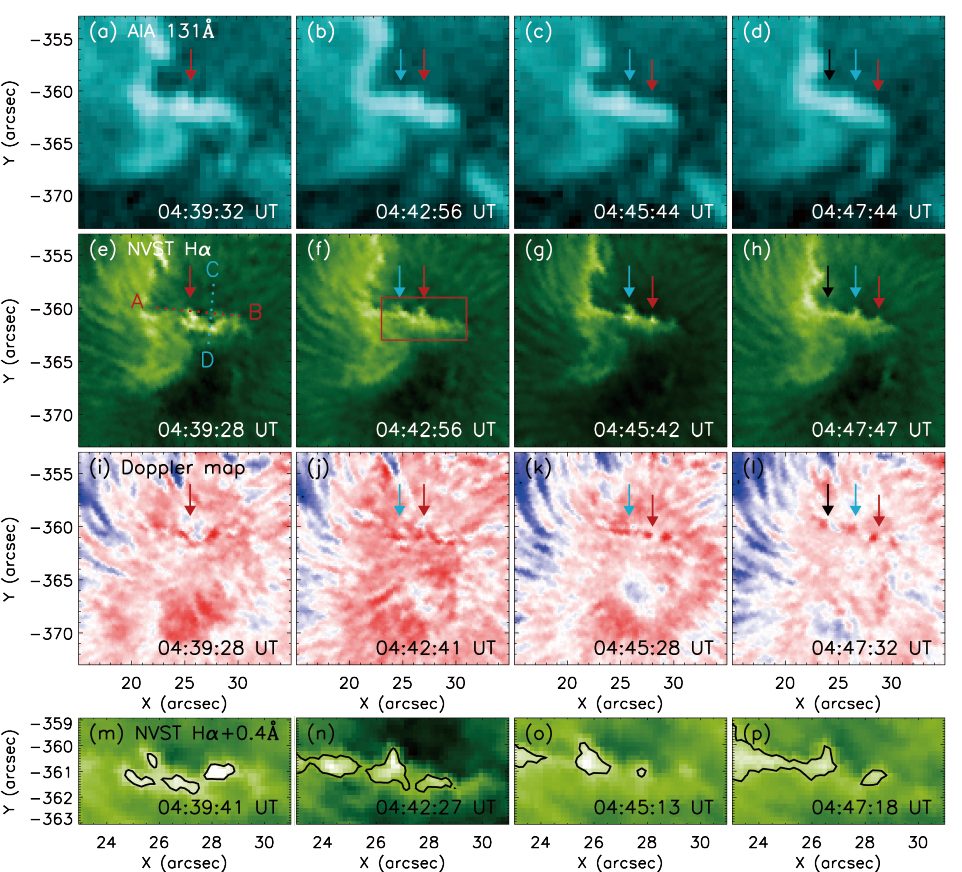}
\caption{
\label{fig:3}Multiwavelength observations of the structures of burrs along the flare ribbon. Panels (a)–(d): SDO/AIA 131 \AA\  image sequence showing the appearance and motion of burrs in the hot corona. Panels (e)–(h): NVST H$\alpha$ image sequence showing the corresponding burr structures in the chromosphere;  the red box in panel (f) indicates the FOV of panels (m)–(p). Panels (i)–(l): H$\alpha$ Doppler proxy map sequence revealing the Doppler proxy signals  at the locations of these structures.  Panels (m)–(p): NVST H$\alpha$ +0.4 \AA\ off-band images of the burrs, where the black contours represent the boundaries defined by the FWHM method. The red, blue, and black arrows point to different burrs appearing at different times to track their movement; note that arrow colors are for feature tracking only and do not represent the sign of the Doppler velocity. The animation of this figure includes AIA 131 \AA, NVST H$\alpha$, and Doppler proxy map images from 04:37 to 05:25 UT with a video duration of 4 s.}
\end{figure*}

\section{Result} \label{sec:Result}

On 2021 April 22, a C3.9-class flare occurred in NOAA AR 12816 (Figure~\ref{fig:1}). The SDO/HMI magnetogram reveals that the flare consisted of an L-shaped negative-polarity ribbon (NR, extending northward) and a J-shaped positive-polarity ribbon (SR, curving southward), with one footpoint of the SR being rooted inside the main sunspot (Figure~\ref{fig:1}(a)). In the EUV, the SDO/AIA 171 \AA\ image (Figure~\ref{fig:1}(b)) shows numerous pre-existing, curved coronal loops that brightened significantly during the flare. Observations in the AIA 304 \AA\ passband (Figure~\ref{fig:1}(c)) clearly delineate the flare ribbons in the chromosphere and transition region. Following the flare's onset, the NR and SR expanded northward and southward, respectively, and formed a classic post-flare loop arcade (Figure~\ref{fig:1}(f)), which is a typical evolutionary feature of two-ribbon flares. Ground-based observations from the NVST provide high-resolution details. The TiO image (Figure~\ref{fig:1}(d)) displays the fine optical structure of the main sunspot and the light bridge background in the region of the SR footpoint, while the H$\alpha$ image (Figure~\ref{fig:1}(e)) shows the corresponding structure of the flare ribbon in the chromosphere.

\begin{figure*}[t!]
\includegraphics[trim={0cm 0cm 0cm 0cm}, clip, width=\textwidth]{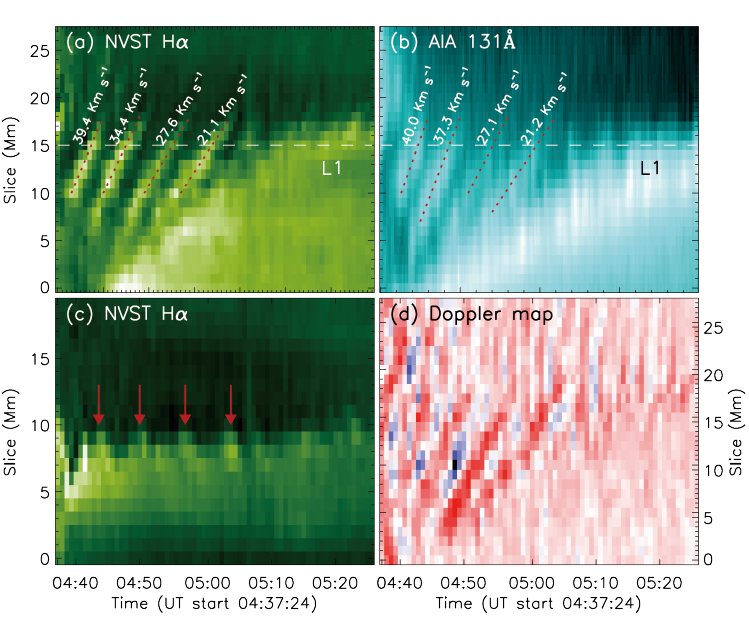}
\caption{
\label{fig:4}Kinematic analysis of the slipping motion of the burrs using time-distance diagrams. Panels (a) and (b): Time-distance diagrams made from the NVST H$\alpha$ and AIA 131 \AA\  data, respectively, along the slit ``A–B" marked in Figure~\ref{fig:3}(e).The slanted bright stripes in these diagrams illustrate the slipping motion of the burrs, and the red dashed lines indicate the measurements of the apparent velocity. Panel (c) is a time-distance diagram made from H$\alpha$ data along the slit ``C–D" in Figure~\ref{fig:3}(e), showing the intermittent eruptions of the burrs as indicated by the red arrows. Panel (d): The corresponding time-distance diagram of the Doppler map, displaying the Doppler proxy signals distribution at the locations of the bright stripes.}
\end{figure*}

Figure~\ref{fig:2} details the evolutionary process of the southern flare ribbon's footpoint, shown as a time series of AIA 171 \AA\  images co-aligned with HMI continuum intensity images. In the initial phase of the flare (Figures~\ref{fig:2}(a) and (b), 04:14:10 UT to 04:21:10 UT), an arc-shaped brightening first appears on the northern side of the sunspot, which may be a precursor brightening in the transition region or low corona. As the flare proceeds, the morphology of this bright region continues to evolve (Figures~\ref{fig:2}(c)–(e)). By 04:40:58 UT (Figure~\ref{fig:2}(f)), a key part of this ribbon footpoint is clearly anchored in the light bridge structure within the sunspot umbra. This location is spatially co-aligned with the chromospheric bright ribbon observed in the NVST H$\alpha$ passband (Figure~\ref{fig:1}(e)), thus establishing a direct physical link between the coronal energy release site and the chromospheric fine structure. This observation provides a precise contextual localization for the subsequent high-resolution analysis of the flare's fine structures, indicating that a significant fraction of the energy release occurred above the light bridge region.

Figure~\ref{fig:3} provides a detailed, multiwavelength study of the evolution of the burr structures. Around 04:39:32 UT, a distinct brightening appears in the AIA 131 \AA\  images (indicated by the red arrow in Figure~\ref{fig:3}(a)), while at the same time, an outwardly protruding burr  structure is observed at the same location in the NVST H$\alpha$ images (red arrow in Figure~\ref{fig:3}(e)). Over the next few minutes (Figures~\ref{fig:3}(b)–(d) and (f)–(h)), we observe that the initial brightening is displaced to the right along the flare ribbon, while a new burr appears successively in its vicinity (indicated by the blue and black arrows). The entire process manifests as the intermittent appearance of burrs at localized points, which subsequently slip along the flare ribbon. This indicates a synchronous response to the energy release process in both the chromosphere and the hot corona. Concurrently, the Doppler proxy maps (Figures~\ref{fig:3}(i)–(l)) show that the brightened locations indicated by the arrows are dominated by strong redshift signals. Faint blueshifts are also discernible in the regions between the redshifts (Figures~\ref{fig:3}(i)–(l)), resulting in an overall pattern of alternating redshifts and blueshifts. 

To accurately characterize the spatial dimensions of the burr structures, we utilized the NVST H$\alpha$ red-wing (+0.4 \AA) intensity images (Figures~\ref{fig:3}(m)–(p)), where these features are distinctly prominent. We applied the FWHM thresholding method to these images to delineate the boundaries of the structures. The analysis reveals that these burr structures are compact, with equivalent diameters of 233--895 km ($0\farcs32$--$1\farcs23$). They exhibit a discrete spatial distribution along the flare ribbon, with separations between adjacent structures of 1129--1739 km ($1\farcs56$--$2\farcs40$). Furthermore, due to the dynamic morphological evolution of the tail features, we characterized their dimensions through manual identification. The measurements yield projected lengths of 700--1370 km ($0\farcs97$--$1\farcs89$) and widths of 310--600 km ($0\farcs43$--$0\farcs83$)

\begin{figure*}[t!]
\includegraphics[trim={0cm 0cm 0cm 0cm}, clip, width=\textwidth]{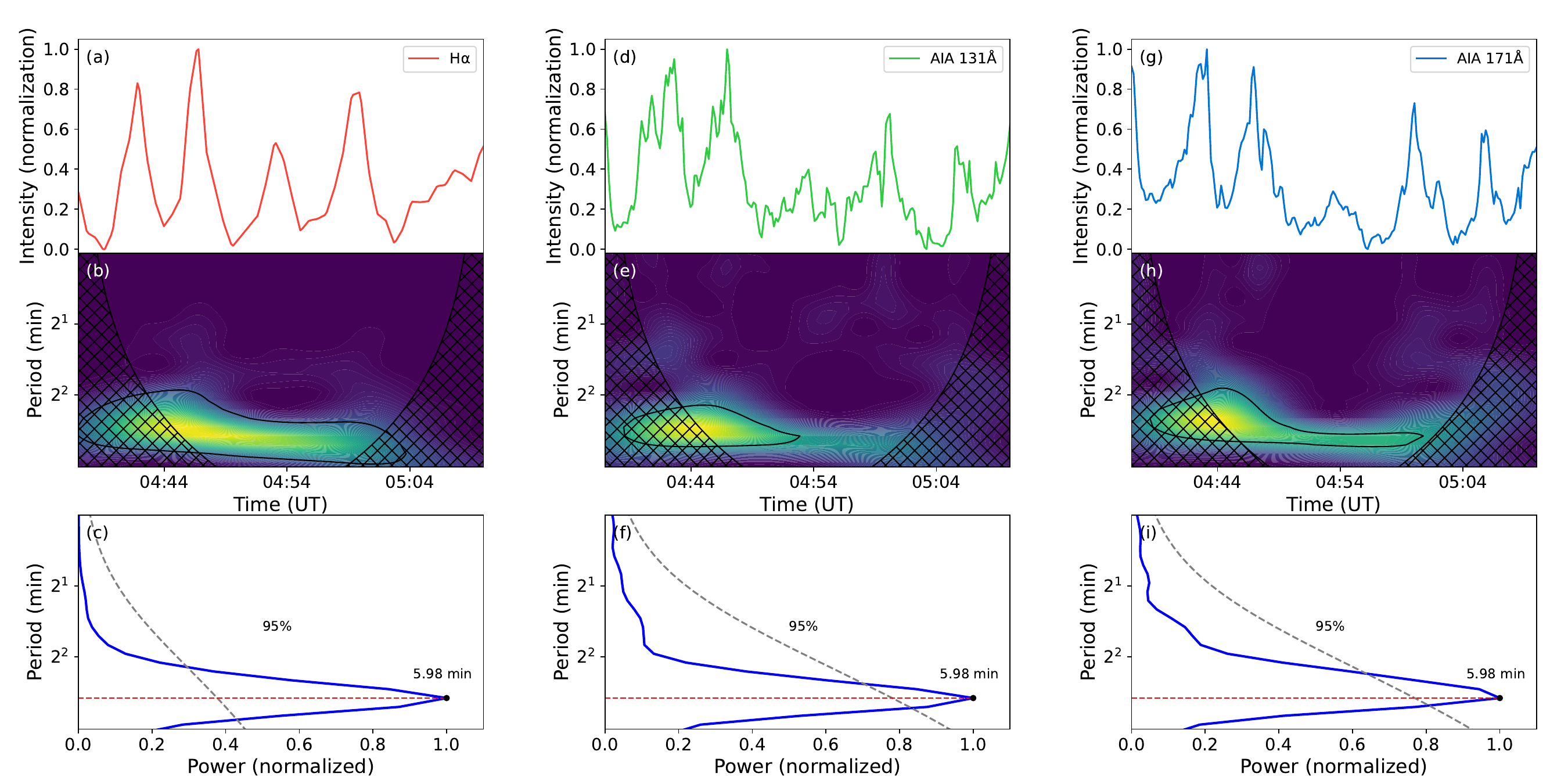}
\caption{
\label{fig:5} Wavelet analysis of the H$\alpha$, AIA 131 \AA, and AIA 171 \AA \ light curves. Panel (a): The H$\alpha$ light curve extracted from the horizontal slice ``L1" in Figure 4(a). Panel (b): The wavelet power spectrum of the light curve in panel (a) from 04:37 to 05:10 UT. The black contour outlines the 95\% significance level. Panel (c): The corresponding global wavelet spectrum, showing a dominant period of $\sim $6 minutes, which is above the 95\% confidence level (dashed gray line). Panels (d)–(f): Similar to panels (a)–(c), but for the AIA 131 \AA \ light curve from slice ``L1" in Figure 4(b). Panels (g)–(i): Similar to panels (a)–(c), but for the AIA 171 \AA \ light curve extracted from the same slice ``L1", which also reveals a consistent dominant period of $\sim $6 minutes.}
\end{figure*}

To investigate the kinematics of the burrs, we constructed time-distance diagrams along the curved slit ``A-B" (marked in Figure~\ref{fig:3}(e)), utilizing both NVST H$\alpha$ and SDO/AIA 131 \AA\ observations (Figures~\ref{fig:4}(a) and (b)). These diagrams reveal a series of discrete, inclined bright ridges, which trace the continuous slipping motion of the burr structures along the flare ribbon. The apparent slipping velocities were determined from the slopes of these bright ridges in the time-distance diagrams. The measurements indicate a distinct deceleration trend, with the velocity decreasing from an initial value of about 40--21 km s$^{-1}$ in the later phase. This kinematic evolution is consistent across both the chromospheric (H$\alpha$) and coronal (AIA 131 \AA) passbands, suggesting a magnetically coupled motion of the footpoints through the atmosphere. Complementing the intensity data, the time-distance diagram constructed from the H$\alpha$ Doppler proxy maps (Figure~\ref{fig:4}(d)) elucidates the spectral characteristics of these moving features. The trajectories of the bright burrs are spatially co-located with strong redshift signals, consistent with chromospheric condensation. Notably, faint blueshift signatures are intermittently observed in the regions trailing the bright ridges or within the gaps between them, forming a spatially alternating Doppler pattern. Additionally, the temporal evolution along the perpendicular slit ``C-D" (Figure~\ref{fig:4}(c)) confirms the recurrent nature of these structures between 04:40 UT and 05:05 UT.

\begin{figure*}[t!]
\includegraphics[trim={0cm 0cm 0cm 0cm}, clip, width=\textwidth]{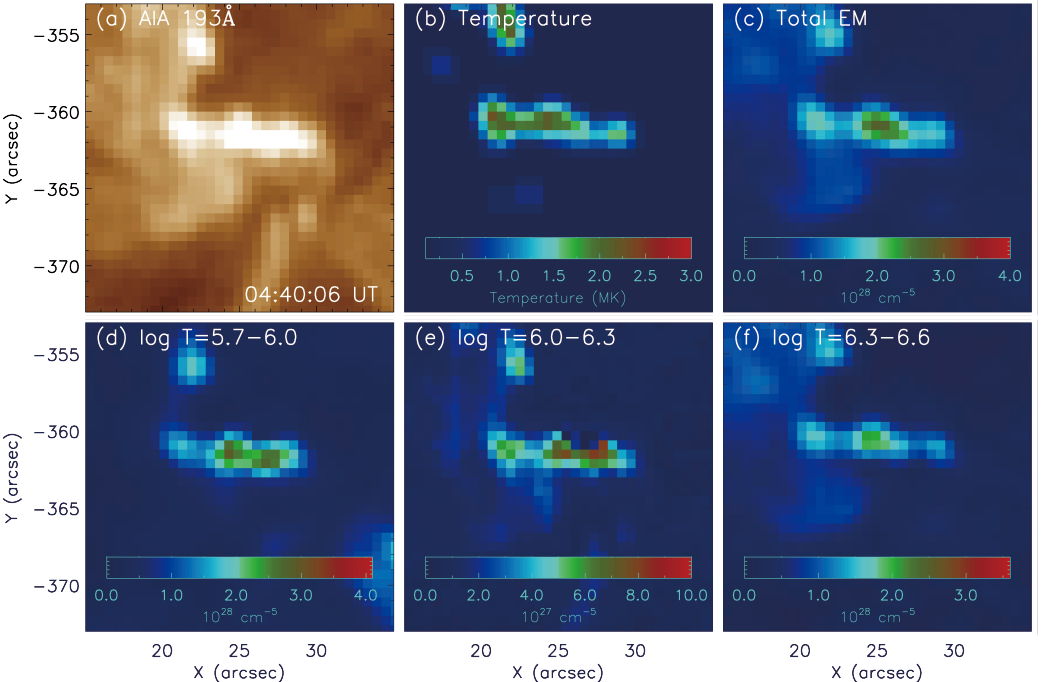}
\caption{
\label{fig:6} Thermodynamic diagnostics of the burr structures. This figure presents the results of a DEM analysis of the burr structures at 04:40:10 UT. Panel (a): AIA 193 \AA \  image for reference. Panel (b): temperature map derived from the DEM analysis, showing the distribution of high-temperature regions corresponding to the bright areas in panel (a). Panel (c): total EM map, showing the high-density regions of the structure. Panels (d)--(f): EM maps for three different temperature ranges, log10 T = 5.7–6.0, 6.0–6.3, and 6.3–6.6, respectively, resolving the structures' different temperature components.}
\end{figure*}

To rigorously determine the characteristic timescales of these recurrent events, we performed a wavelet analysis on the normalized intensity light curves extracted from the horizontal slice ``L1" (indicated in Figures~\ref{fig:4}(a) and (b)). The light curves obtained from NVST H$\alpha$, AIA 131 \AA, and AIA 171 \AA\ (Figures~\ref{fig:5}(a), (d), and (g)) all exhibit pronounced quasiperiodic pulsations. The corresponding wavelet power spectra (Figures~\ref{fig:5}(b), (e), and (h)) display enhanced power localized around a specific period throughout the event duration. Integration of these signals into global wavelet power spectra (Figures~\ref{fig:5}(c), (f), and (i)) reveals a dominant periodicity of $\sim$6 minutes, which is statistically significant above the 95\%  confidence level.

Finally, to diagnose the detailed thermodynamic properties of the burr structures at 04:40:10 UT, we performed a DEM analysis using AIA data (Figure~\ref{fig:6}). The temperature map (Figure~\ref{fig:6}(b)) reveals that the bright regions in the AIA 193 \AA\  reference image (Figure~\ref{fig:6}(a)) are hot, while the total EM map (Figure~\ref{fig:6}(c)) indicates a high emission measure. An examination of the EM distribution across different temperature bins shows some contribution from cooler plasma ($\log_{10} T$= 5.7--6.0; Figure~\ref{fig:6}(d)), a peak contribution in the intermediate range ($\log_{10} T$ = 6.0--6.3; Figure~\ref{fig:6}(e)) whose morphology is consistent with AIA 193 \AA. In the higher-temperature range ($\log_{10} T = 6.3$--$6.6$, $\sim$2--4\ $\mathrm{MK}$; Figure~\ref{fig:6}(f)) there is also an emission-measure contribution, but its intensity is significantly lower than the peak interval at $1$--$2\ \mathrm{MK}$. These analyses indicate that the bulk of the plasma composing the burrs is concentrated at $1$--$2\ \mathrm{MK}$. This temperature range closely matches the response-function peak regions of the AIA 193 \AA\ channel, which is sensitive to $\sim$1.5\ $\mathrm{MK}$ ($\log_{10} T \approx 6.1$--$6.2$) and $\sim$2.0\ $\mathrm{MK}$ ($\log_{10} T \approx 6.3$).

\section{Discussion and summary} \label{sec:Discussion}

 In this study, we present a detailed analysis of the fine structures of the flare ribbon anchored at a sunspot light bridge during a C3.9-class solar flare, utilizing high-resolution multiwavelength observations. Our observations reveal discrete substructures appearing on the macroscopically continuous flare ribbon, characterized by main dot-like cores with  tail features, which we term as burrs. These features exhibit distinct periodic slipping motions and characteristic Doppler proxy signatures. This work provides a comprehensive observational study linking the morphological evolution and kinematics of these fine structures, offering observational evidence that identifies them as elementary units of flare energy release anchored at a sunspot light bridge.

Morphologically, the burr structures resemble the riblets  \citep{Thoen_2026,Singh_2025}, sharing streak-like appearances and comparable geometries. With projected lengths of 700--1370 km and widths of 310--600 km near the disk center imply that these structures are highly inclined with respect to the local vertical. However, unlike riblets distributed uniformly along the ribbon front, the burrs are spatially confined to the ribbon hook region. Their equivalent diameters (233--895 km) and inter-core separations (1129--1739 km) exceed the fine-scale structures in \cite{Thoen_2025a} and \cite{Yadav_2025}, likely due to broader emission in the H$\alpha$ +0.4 \AA\  passband and boundary uncertainties from irregular morphology. Notably, the inter-core separation is comparable to the $\sim$1.7--1.9 Mm dominant mode in \citet{French_2025}, supporting a tearing-mode instability origin.

Given their morphology and confinement to the hook region, we interpret these structures as plasmoids (magnetic flux ropes) generated by tearing-mode instability in a fragmenting current sheet, rather than Kelvin-Helmholtz instability  \citep{Ofman_2011} or velocity shears \citep{Parker_2017}. Theoretical models predict such instabilities produce small-scale flux ropes that impact the lower atmosphere as localized fine structures \citep{Wyper_2021,Yan_2022}. Here, energetic particle injection forms bright footpoint cores, while the tails represent the chromospheric mapping of the oblique, twisted magnetic field. This corroborates the role of tearing-mode instability in fragmenting flare energy release, consistent with recent high-resolution studies \citep{Thoen_2025a,Yadav_2025}.

The Doppler proxy map in Figure~\ref{fig:3}(d) reveals faint blueshifts in gaps or trailing redshifted cores. Reconnection drives thermal and non-thermal injection along these lines, triggering chromospheric condensation \citep{Ichimoto_1984,Fisher_1985,Canfield_1990,Milligan_2009} visible as redshifted bright cores \citep{Thoen_2025a}. DEM analysis confirms these cores as primary energy deposition sites dominated by 1–2 MK plasma (Figure~\ref{fig:6}), consistent with the stationary bright points in \citep{Yadav_2025}.  The faint chromospheric blueshifts detected in high-resolution observations are often interpreted as ``gentle evaporation" \citep{Fisher_1985,Milligan_2006}, ``chromospheric bubbles" lifted by hot plasma \citep{Tei_2018}, or coronal rain ``splash-back" \citep{Pietrow_2024}. In contrast, the blueshifts here are strictly localized to the tail feature, vanishing simultaneously with it. This specific spatiotemporal correlation rules out general evaporation, pointing instead to intrinsic tail dynamics. Following \citet{Wyper_2021}, we interpret the tail as the chromospheric projection of an oblique magnetic flux rope; its upward ejection, combined with magnetic untwisting and volumetric expansion, drags plasma outward, naturally generating the observed blueshift.

 Observations reveal that the slipping kinematics of the burrs exhibit a distinct deceleration trend, with their propagation speed along the flare ribbon decreasing from about 40--21 km s$^{-1}$ (Figure~\ref{fig:3}). Theoretically, the slipping velocity of magnetic footpoints is modulated by the squashing factor ($Q$) of the QSL \citep{Aulanier_2006,Janvier_2013}.  The initial rapid motion and distinct morphology suggest that reconnection occurs at the high-$Q$ core of the QSL, namely the hyperbolic flux tube (HFT), which amplifies the apparent motion of the footpoints. The subsequent deceleration and structural fading are accompanied by the outward expansion of the flare ribbons, implying a geometric drift of the reconnection footpoints from the high-$Q$ core toward the low-$Q$ periphery. Furthermore, this deceleration persists throughout the gradual phase, consistent with the expected decline in the global reconnection rate. This kinematic evolution is likely a result of the combined effects of the geometric transition and the decay of the global reconnection rate during the gradual phase. Notably, this observational trend of significant velocity reduction during the gradual phase is consistent with the results reported by \citet{Li_2015}.

Multiwavelength observations reveal that the discrete burr structures distributed along the flare ribbon exhibit quasiperiodic slipping features with a period of $\sim$6 minutes (Figure~\ref{fig:5}). This timescale falls within the typical range of solar flare quasiperiodic pulsations (QPPs) \citep{Nakariakov_2009,Van_2016} and is also close to the 3–5 minute period of solar photospheric $p$-mode oscillations \citep{Chen_2006}. First, external $p$-mode coupling offers a possible explanation, namely that upward-propagating MHD waves modulate the magnetic reconnection rate \citep{Chen_2006,Li_2015}. Second, theoretical studies indicate that the generation, coalescence, and dynamic interaction of plasmoids \citep{Biskamp_1986,Shibata_2001} are inherently quasiperiodic processes \citep{Kliem_2000}. The energy release characteristics triggered by these processes span a wide temporal range, covering from tens of seconds to tens of minutes \citep{French_2025}. Therefore, rather than identifying a single driver, we suggest that the observed periodicity likely reflects a coupling of these mechanisms: external $p$-mode oscillations may act as a perturbation source that triggers or modulates the tearing-mode instability within a marginally stable current sheet.

An interesting question arising from this study is the role of the light bridge in this process. Figure~\ref{fig:2} clearly shows  that the burrs are anchored in the light bridge region of the primary sunspot, indicating that the light bridge plays a  crucial role in flare energy release. First, acting as a trigger site, the high magnetic shear environment of the light bridge provides ideal conditions for recurrent magnetic reconnection \citep{Toriumi_2015a,Toriumi_2015b}. Second, the light bridge acts as a confining channel that typically corresponds to the photospheric base of a QSL \citep{Demoulin_1996, Zhao_2016}, where the magnetic topology naturally constrains the slipping path of the reconnected field lines. Consequently, the light bridge not only localizes the energy release, but its complex underlying magnetic boundary also acts as a core factor driving and constraining the entire dynamic process, strongly supporting the view that complex photospheric magnetic boundaries provide critical constraints on overlying energy-dissipation patterns \citep{Pietrow_2024}.

Although this study provides comprehensive diagnostics, certain limitations remain. First, the analysis based on a single event needs to be extended to a broad statistical study to verify the universality of this light bridge-confined, tearing-mode-driven mechanism and its dependence on flare properties. Second, the current discussion regarding magnetic topology is primarily based on indirect inferences. Future work should incorporate NLFFF modeling \citep{Wiegelmann_2004,Sun_2012} to reconstruct the 3D magnetic field, thereby precisely locating the QSL and quantifying the magnetic shear in the light bridge region. Finally, constrained by current resolution, the tearing process within the current sheet remains unresolved. Higher-resolution instruments, such as DKIST , are expected to reveal finer structures within the burrs, thus elucidating the fundamental units of flare energy release.

 In summary, our observations provide strong evidence that flare ribbons are composed of discrete elementary units driven by fragmented magnetic reconnection. Specifically, despite the reliance on Doppler proxies and the analysis of a single event, the spatial association with light bridge anchoring, coupled with the morphological and spectral evidence of fragmentation, strongly supports the interpretation of these fine-scale structures as plasmoids resulting from tearing-mode instability in the reconnection current sheet. Furthermore, the observed $\sim $6 minute periodicity implies a potential modulation of the tearing-mode instability by photospheric $p$-mode oscillations, highlighting the complex coupling between the lower atmosphere and coronal energy release processes. Ultimately, these findings underscore the necessity of future statistical investigations and higher-resolution campaigns, such as those utilizing the DKIST, to fully resolve the internal dynamics of these fine-scale structures and validate the proposed magnetic topology through advanced 3D modeling.

\begin{acknowledgments}

The authors are grateful to the anonymous referee for valuable suggestions. We also thank the NVST and SDO science teams for providing the data. This work is sponsored by the National Key R\&D Program of China, under No. 2024YFA1612001; the Strategic Priority Research Program of the Chinese Academy of Sciences, grant No. XDB0560000; the National Science Foundation of China (NSFC), under Nos. 12325303, 12473059, 12373115, 12203097, 11973084, 12127901, and 12403065; the Yunnan Key Laboratory of Solar Physics and Space Science, under No. 202205AG070009; and the Yunnan Fundamental Research Projects, under Nos. 202301AT070347 and 202301AT070349. 

\end{acknowledgments}

\bibliography{reference}{}
\bibliographystyle{aasjournalv7}



\end{document}